%% file: main.tex
\documentclass{article} 
\usepackage{iclr2026_conference,times}

\input{math_commands.tex}

\usepackage[utf8]{inputenc} 
\usepackage[T1]{fontenc}    
\usepackage{hyperref}       
\usepackage{url}            
\usepackage{booktabs}       
\usepackage{amsfonts}       
\usepackage{nicefrac}       
\usepackage{microtype}      
\usepackage{xcolor}         
\usepackage{comment}
\usepackage{hyperref}
\usepackage{hyperref}
\usepackage{url}
\usepackage{caption}
\usepackage{subcaption}
\usepackage{graphicx}
\usepackage{float}
\usepackage{amsmath}
\usepackage{tabularx,booktabs}
\usepackage{wrapfig}

\definecolor{purple}{RGB}{147, 112, 219}
\newcommand{\cathy}[1]{\textcolor{purple}{\bf\small [#1 --cathy]}}

\title{Efficient Dataset Selection for Continual Adaptation of Generative Recommenders}

\author{
\begin{minipage}{\textwidth}
\centering
\normalfont
\vspace{0.3in}
Cathy Jiao$^{2*}$ \quad Juan Elenter$^{1}$ \quad Praveen Ravichandran$^{1}$ \quad Bernd Huber$^{1}$\\
Joseph Cauteruccio$^{1}$ \quad Todd Wasson \quad Timothy Heath$^{1}$ \quad Chenyan Xiong$^{2}$\\
Mounia Lalmas$^{1}$ \quad Paul Bennett$^{1}$\\[4pt]
$^{1}$Spotify \quad $^{2}$Carnegie Mellon University
\end{minipage}
}

\iclrfinalcopy 
\begin{document}

\maketitle

\begin{abstract}
Recommendation systems must continuously adapt to evolving user behavior, yet the volume of data generated in large-scale streaming environments makes frequent full retraining impractical. This work investigates how targeted data selection can mitigate performance degradation caused by temporal distributional drift while maintaining scalability. We evaluate a range of representation choices and sampling strategies for curating small but informative subsets of user interaction data. Our results demonstrate that gradient-based representations, coupled with distribution-matching, improve downstream model performance, achieving training efficiency gains while preserving robustness to drift. These findings highlight data curation as a practical mechanism for scalable monitoring and adaptive model updates in production-scale recommendation systems.
\end{abstract}

\input{sections/introduction}

\input{sections/datasets}

\input{sections/methods}

\input{sections/experiment_setup}

\input{sections/results}
\input{sections/conclusion}

\newpage

\bibliography{iclr2026_conference}
\bibliographystyle{iclr2026_conference}

\appendix
\input{sections/appendix}
\end{document}

%% file: math_commands.tex
\usepackage{amsmath,amsfonts,bm}

\def\eqref#1{equation~\ref{#1}}

\def\1{\bm{1}}

\DeclareMathAlphabet{\mathsfit}{\encodingdefault}{\sfdefault}{m}{sl}
\SetMathAlphabet{\mathsfit}{bold}{\encodingdefault}{\sfdefault}{bx}{n}



%% file: sections/introduction.tex
\section{Introduction}\label{sec:introduction}

\begingroup
\renewcommand\thefootnote{}
\footnotetext{* Work done while interning at Spotify.

Corresponding author: \texttt{juane@spotify.com}}
\endgroup

Sequential generative recommendation models take sequences of user interaction histories and output a distribution over likely next interactions. Motivated by recent advances in sequence modeling with transformer architectures, a growing line of work has proposed several sequential recommenders, including Meta’s HSTU (\cite{zhai2024actions}), SASRec (\cite{kang2018self}), and TallRec (\cite{Bao_2023}). While these models can capture rich temporal dependencies in user behavior, deploying them in real streaming settings introduces a central challenge: user behavior is not stationary, and the data stream is effectively unbounded.

User behavior is known to exhibit behavioral drift, where the mapping from historical interactions to future preferences changes over time. In music and podcast consumption, drift may arise from seasonality, shifts in behavioral context (e.g., commuting versus at-home listening), and catalog changes such as new releases. Drift may also reflect longer term preference evolution, as users change life activities that influence their streaming behavior or develop new genre affinities. Consequently, interactions collected in the more distant past may become less predictive of near-term behavior, and models trained on stale data can experience systematic performance degradation.

This work frames the resulting system pressure along two axes. First, temporal adaptation: if we do not update a deployed sequential recommender, performance degrades over time as drift accumulates (Figure \ref{fig:setup}). Second, data scaling under constraints: continuously receiving new interactions suggests an opportunity to benefit from scaling laws by training on ever more data. However, the cumulative scale of the stream precludes repeatedly training on all available data. As user histories extend over days, months, and years, na\"ively incorporating all observed interactions becomes increasingly impractical in terms of storage, compute, and retraining latency. The practical question is therefore not only whether to update, but how to update efficiently in a manner that captures the benefit of additional data while operating within compute budgets.

To address these issues, we study data selection methods for curating compact, high-quality training sets that support continual updates while respecting compute budgets. As a motivating baseline, Figure \ref{fig:setup} suggests that even random selection can be effective: retraining every six months adding 20\% of newly observed data yields performance that improves substantially over no updates, yet remains below training on all new data. This gap motivates more principled selection strategies that aim to preserve and amplify the benefit of additional data while using only a fraction of the stream.

Building on recent work emphasizing the roles of data representation and sampling strategy \citep{pruthi2020estimating, xia2024less, jiao2025datelm}, we explore selection methods across these two components, along with their associated trade-offs. In the remainder of this paper, we show that certain choices of representation (notably gradient-informed signals) can substantially improve selection quality, and that sampling strategies combining distribution matching with diversity can further boost downstream performance.

\textbf{Contribution.} We present an empirical study of continual adaptation for sequential recommenders using a subsample spanning ten years of real-world music and podcast interaction data. We show that temporal drift leads to systematic performance degradation, and that retraining on small, carefully selected subsets of recent user interactions can recover a substantial portion of the gains of full retraining under compute constraints. We further demonstrate that gradient-based representations and sampling strategies combining distribution matching with diversity consistently yield more effective updates in this streaming setting.

\input{figures/monthly_setup}

%% file: figures/monthly_setup.tex
\begin{figure}[ht]
    \centering
    \includegraphics[width=0.7\linewidth]{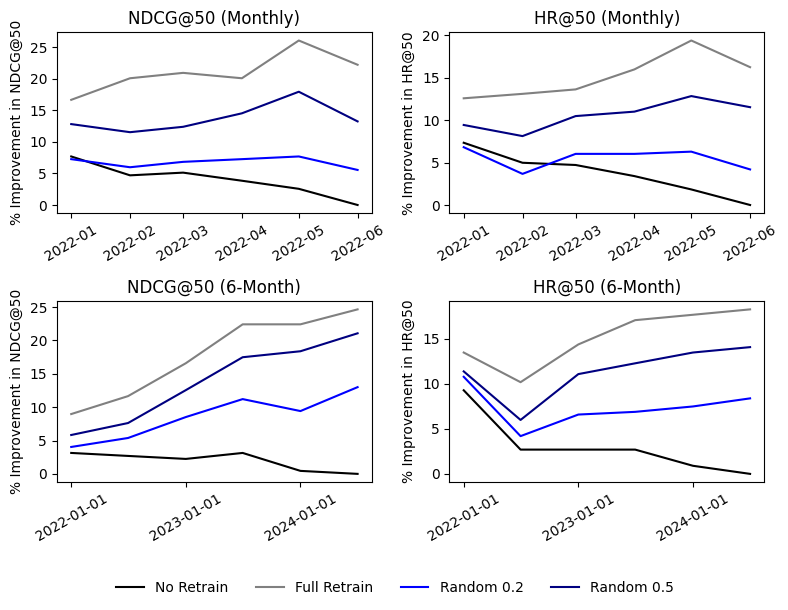}
\caption{
Relative performance improvement of the HSTU model over time, measured with NDCG@50 and HR@50.
The top row reports retraining and evaluation performed monthly, while the bottom row reports results every six months.
At each time window, model updates are performed by continuing training from the previous checkpoint and incorporating previously unseen sequences.
We compare four strategies: (i) no further training, (ii) training with 100\% of new data, (iii) training with a random 20\% subset of new data, and (iv) training with a random 50\% subset of new data.
}

    \label{fig:setup}
\end{figure}

%% file: sections/datasets.tex
\section{Datasets and Models}

In this section, we describe the data and modeling choices in our experiments: a decade-scale music and podcast streaming corpus with evolving users and item sets (Section \ref{sec:dataset}), and a sequential generative recommender model suited to non-stationary user historical sequences (Section \ref{sec:models}).

\subsection{Dataset}\label{sec:dataset}

We use a longitudinal sample from a proprietary music and podcast streaming dataset spanning 2015–2025. The dataset contains on the order of 10K users and a vocabulary on the order of 10M items (tracks and podcasts).

Each user history is a time-ordered sequence of interaction events. An interaction event is represented by an item and an action  element, where the item is an integer identifier and the action is a tuple of categorical fields describing how the interaction occurred. In this work, we use two action fields: (i) reason end, which indicates why playback ended, and (ii) interaction type, which describes the context of the stream. Some examples of reason end include the user skipping a track, pausing playback, or exiting the application. Examples of interaction type include streams initiated from a user-curated playlist or directly via a catalog search. At each time step, an interaction is represented as $(o_t, a_t)$ where $o_t$ is the item id and $a_t = (\mathrm{reason\_end}_t, \mathrm{interaction\_type}_t)$.

In streaming settings, new user histories and new items arrive continuously over time. In this dataset, both the number of newly observed user sequences and the number of newly introduced items grow steadily from month to month. This gradual increase reflects a realistic non-stationary environment and makes the dataset well suited for studying continual adaptation under temporal drift.

\subsection{Models}\label{sec:models}

Following recent works in sequential generative recommender models, we use Meta's HSTU \citep{zhai2024actions}. This model takes a user’s listening history as a sequence and is trained autoregressively to predict the next interaction, which may correspond to either an item or an action. We use separate output heads for item and action prediction, both trained with cross-entropy loss, with sampling applied only to the item prediction objective. While we primarily used HSTU for experiments, it is possible to expand this to other sequential generative models, such as SASRec \citep{kang2018self}. 

\begin{figure}[h] 
    \centering
    \includegraphics[width=0.49\textwidth]{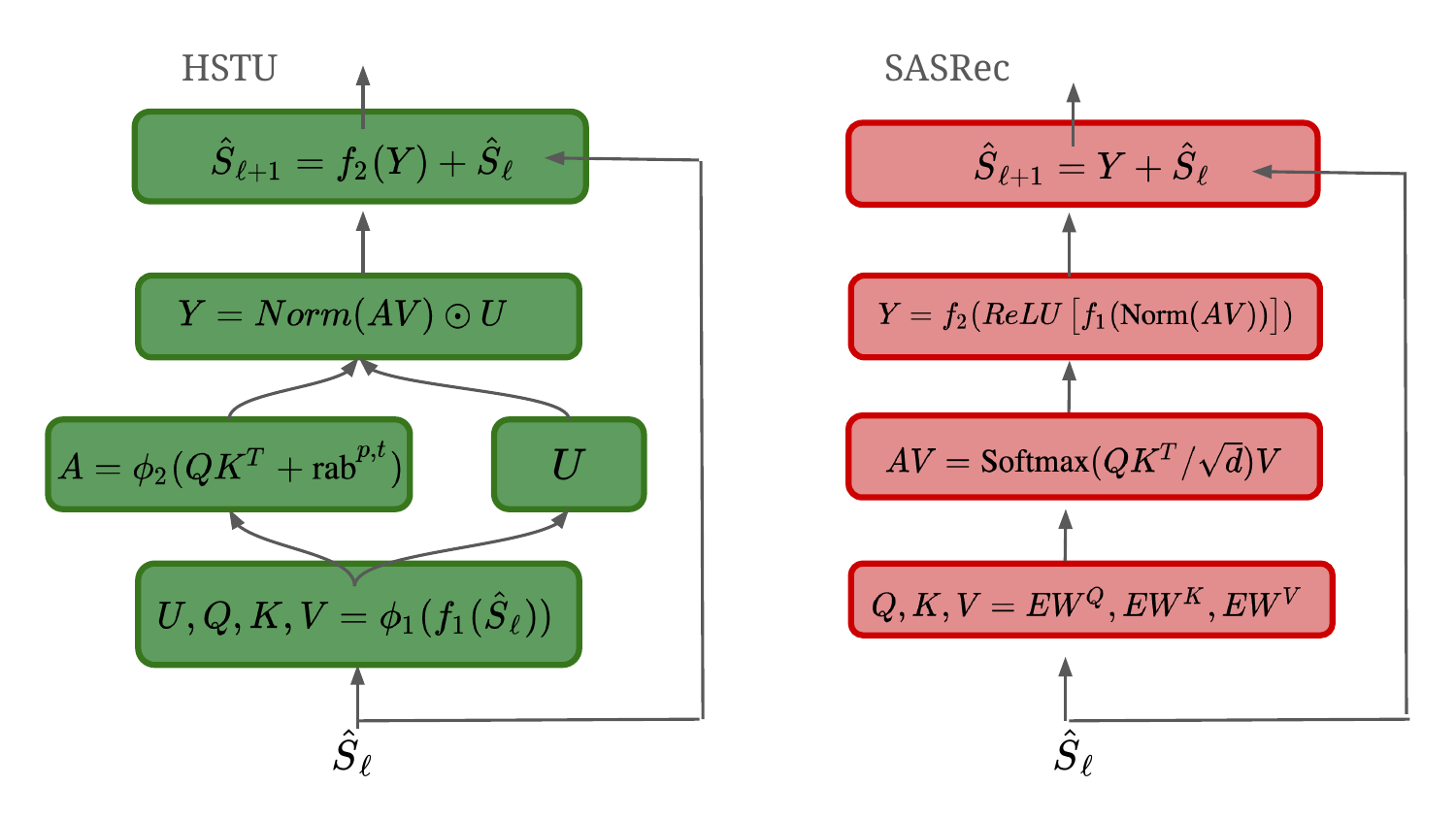}
    \caption{Diagram of the main block in HSTU and comparison with SASRec}
    \label{fig:hstusasrec}
\end{figure}

HSTU \citep{zhai2024actions} introduces a modified attention mechanism optimized for non-stationary streaming data (see Figure \ref{fig:hstusasrec}). The architecture consists of a cascade of blocks, each with three components: projection, spatial aggregation, and pointwise transformation. These components operate as follows:
\begin{align}
& U, V, Q, K = \text{Split}(\phi_1(f_1(\mathbf{E}))) \label{split} \\
& \text{Attention}(Q, K, V) = AV = \phi_2\left(Q  K^T + \text{rab}^{p,t}\right)V \label{att} \\
& \hat{\mathbf{S}} = f_2(\text{Norm}(A  V ) \odot U) \label{feat}
\end{align}
where $f_i(X) = W_iX + b_i$ are MLPs, $\phi_1$ and $\phi_2$ represent the SiLU nonlinear activation function~\citep{elfwing2018sigmoid}, $\text{rab}^{p,t}$ denotes relative attention biases, incorporating both positional and temporal information~\citep{raffel}, $\odot$ represents element-wise multiplication and $\text{Norm}$ refers to layer normalization for improved training stability. 

HSTU introduces three main innovations that distinguish it from previous transformer-based architectures for recommendation. Rather than relying solely on absolute positional information, HSTU incorporates the time difference between pairs of sequence elements to compute attention weights \citep{raffel}. Another distinction is its adoption of the SiLU activation function in place of softmax for computing attention weights (see equation~\ref{att}). This choice is motivated by the fact that the number of prior interactions associated with a target item provides an indication of user engagement intensity, which is difficult to capture after applying softmax normalization. Finally, HSTU does feature interaction through a point-wise transformation, utilizing a Hadamard product between the attention-pooled features and the input features (see equation~\ref{feat}).

Overall, HSTU is a strong fit for our streaming dataset due to its ability to model long, non-stationary user histories. Its relative temporal biases and SiLU-based attention preserve engagement signals over long horizons, making it well-suited for large-scale recommendation under continual temporal drift.

%% file: sections/methods.tex
\section{Data Selection Methods}\label{sec:methods}
\input{figures/cont_learning_setup}

In this section, we describe the data selection methods used for our experiments. In particular, given a set of training samples as $\mathcal{D}_{train} = \{x_i\}_{i=1}^n$, our objective is to select a small high-quality subset $\mathcal{D}_{select} \subset \mathcal{D}_{train}$ where $|\mathcal{D}_{select}
| \ll |\mathcal{D}_{train}|$ for training. Following recent works \citep{pruthi2020estimating, xia2024less}, we conduct data selection in three stages. First, we map each user sequence to an intermediate vector representation. Second, we assign a score to each representation that reflects its utility for training. Finally, we construct $\mathcal{D}_{select}$ by sampling training examples according to these scores.

\textbf{User Data:}  In our setting, a training sample $x_i$ is a user interaction history represented as a sequence of (item, action) pairs:
$x_i = {(o_{i1}, a_{i1}), ..., (o_{ir}, a_{ir})}$,
where $o_{ij}$ is the item id and $a_{ij}$ is the associated action (reason end and interaction type).

\textbf{Gathering Representations:} In this stage, we investigate three representation types: token-based, model-based, and gradient-based representations. Below, we describe the representations in detail.

\begin{enumerate}
    \item \textit{Token-based representation}: We map each interaction event $(o, a)$ into a multiset of discrete tokens, including an item token and action tokens (e.g., item:$o$, reason end:$v$, interaction type:$u$). We then represent each history $x$ as a bag of these tokens and compute similarity with BM25.
\item \textit{Model-based representation (RepSim)}: We tokenize each interaction event $(o, a)$ and feed the resulting sequence into the HSTU model $\mathcal{M}_{\theta}$. Let $\{h_1, \ldots, h_r\}$ denote the sequence of hidden states from the final layer of $\mathcal{M}_{\theta}$ for a user history $x$. We obtain a fixed-dimensional representation $rep(x)$ by applying mean pooling over these final-layer hidden states, yielding a single vector representation of the sequence.

\item \textit{Gradient-based representation (GradSim)}: For each history $x_i$, we compute the gradient of the loss for predicting the final item in the sequence, conditioned on the entire preceding (item, action) pair history. Specifically, we define:
\[
\mathrm{rep}(x_i)
= \nabla_{\theta} \mathcal{L}
\bigl(
o_{ir}
\mid
(o_{i1}, a_{i1}), \ldots, (o_{i,r-1}, a_{i,r-1})
\bigr),
\]
and construct the representation by extracting gradients with respect to the parameters of the final attention layer. As in RepSim, we apply mean pooling to obtain a single vector representation.
\end{enumerate}

\textbf{Scoring Representations:} In this stage, we score the training data samples with the aid of a reference set $\mathcal{D}_{ref} = \{x'_j\}_{j=1}^m$ where $m \ll n$. The reference set consists of a small set of samples which represent target behaviour that the model should learn. For instance, in past works the reference set is typically sampled from the validation set of a target task \citep{pruthi2020estimating, xia2024less, yu2024mates, jiao2025datelm}. Since our setting deals with temporal shifts in user listening preferences over time,  we sample $\mathcal{D}_{ref}$ from user listening histories from the most recent month. Section \ref{sec:experimental_setup} contains additional details regarding reference data setup.

We compute a set of scores $\mathcal{S} = \{ s(x) \}_{x \in \mathcal{D}_{\mathrm{train}}}$ for each training sample (i.e., user listening history) with respect to a reference dataset $\mathcal{D}_{\mathrm{ref}}$. Specifically, the score for a given listening history $x_i$ is defined as the average similarity between its representation and the representations of all listening histories in $\mathcal{D}_{\mathrm{ref}}$:
\begin{equation}
s(x)
= \frac{1}{|\mathcal{D}_{\mathrm{ref}}|}
\sum_{x' \in \mathcal{D}_{\mathrm{ref}}}
\operatorname{sim}\!\left(\operatorname{rep}(x), \operatorname{rep}(x')\right).
\end{equation}
where $\operatorname{rep}(\cdot)$ denotes the representation function, and $\operatorname{sim}(\cdot,\cdot)$ denotes a similarity metric. For token-based representations, we use BM25 (\cite{robertson2009probabilistic}) as the similarity metric, and for model/gradient-based representations, we use cosine similarity.

\textbf{Data Sampling:} In the final stage, we construct the selected training dataset
$\mathcal{D}_{\mathrm{select}} \subset \mathcal{D}_{\mathrm{train}}$
using the scores $\mathcal{S} = \{ s(x) \}_{x \in \mathcal{D}_{\mathrm{train}}}$.
We explore two broad classes of sampling strategies: \emph{ranking-based}
and \emph{probabilistic} methods. Specifically, we consider the following:

\begin{enumerate}
    \item \textit{Top-$K$/Bottom-$K$ (Ranking-based).}
    Select the top-$K$, bottom-$K$, or a mixture of top- and bottom-ranked
    training samples according to their scores $s(x)$.

    \item \textit{Weighted (Probabilistic).}
    Sample $K$ training examples from $\mathcal{D}_{\mathrm{train}}$
    with probabilities proportional to their scores $s(x)$.

    \item \textit{KNN-Weighted (Probabilistic).}
    Cluster the training representations into $C$ clusters.
    From each cluster, select $K / C$ samples using weighted sampling
    based on $s(x)$. 

    \item \textit{Diverse-Weighted (Probabilistic).}
    A greedy, iterative sampling strategy inspired by \cite{wang2024greats}.
    At each iteration $t$, we sample a small batch of training samples, $\mathcal{B}_t$, using weighted sampling.
    For each remaining example
    $x \in \mathcal{D}_{\mathrm{train}} \setminus \mathcal{B}_t$,
    we update its score to penalize redundancy: 
    \begin{equation}
        s(x) \leftarrow s(x)
        - \frac{1}{|\mathcal{B}_t|}
        \sum_{x' \in \mathcal{B}_t}
        \operatorname{sim}\!\left(\operatorname{rep}(x), \operatorname{rep}(x')\right).
    \end{equation}
\end{enumerate}

%% file: figures/cont_learning_setup.tex
\begin{figure}[t]
    \begin{subcaptionbox}{Training Stage \label{fig:train}}[0.49\textwidth]
        {\includegraphics[width=\linewidth]{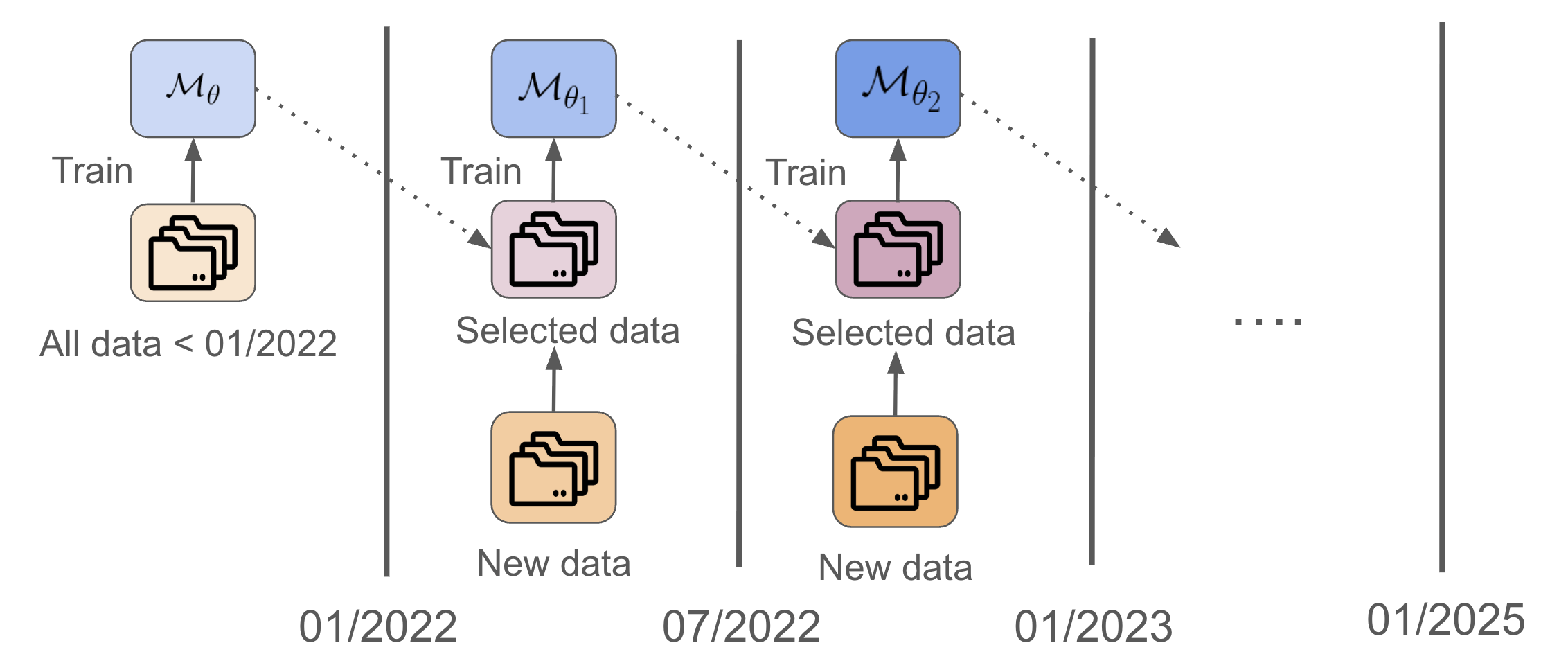}}
    \end{subcaptionbox}
    \begin{subcaptionbox}{Evaluation Stage \label{fig_eval}}[0.49\textwidth]
        {\includegraphics[width=\linewidth]{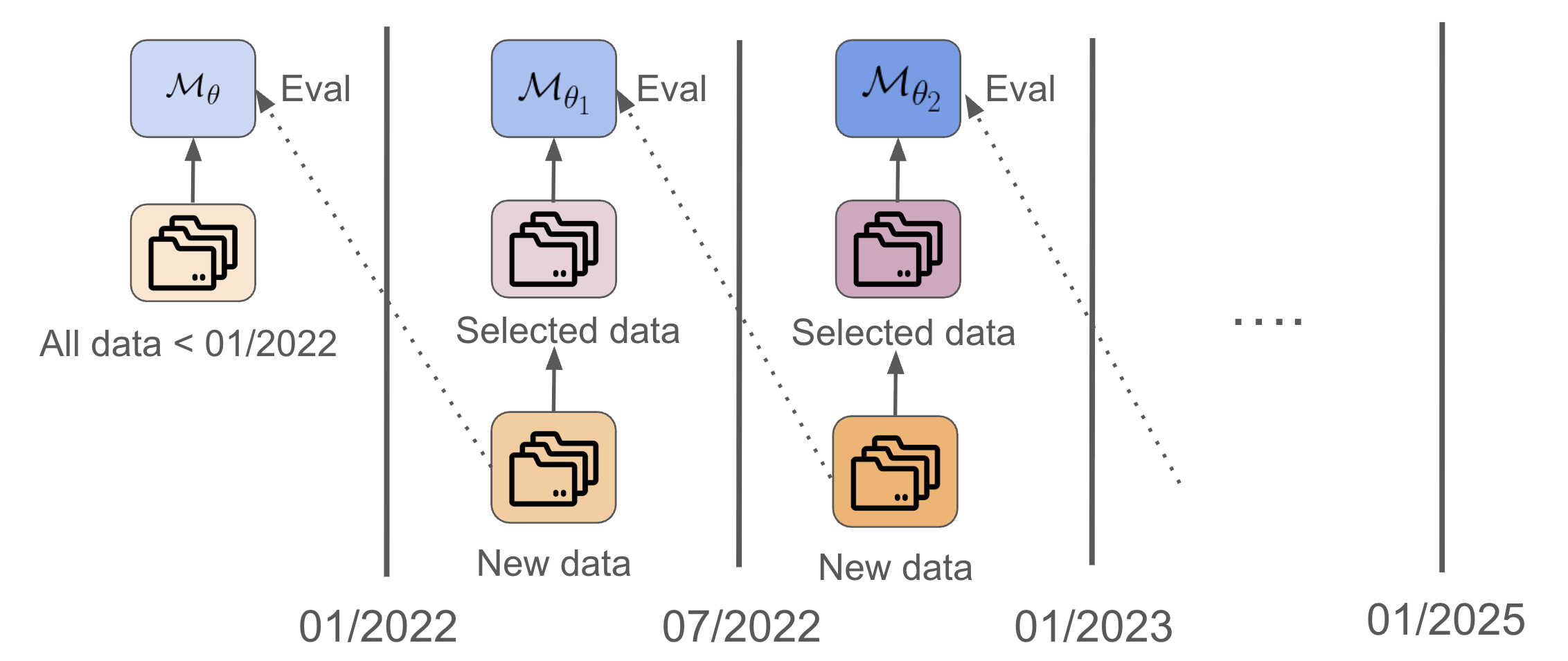}}
    \end{subcaptionbox}
    \caption{Our training and evaluation pipeline in a continuous learning setting. Training is conducted in 6-month intervals.}
    \label{fig:cont_learning_setup}
\end{figure}

%% file: sections/experiment_setup.tex
\section{Experimental Setup}\label{sec:experimental_setup}

We evaluate data selection for continual adaptation using the rolling protocol shown in Figure~\ref{fig:cont_learning_setup}, which provides a natural framework for assessing the effectiveness of data selection methods over time. We first train the HSTU model (Section~3.2) on the proprietary streaming dataset (Section~3.1) from 2015 up to 01/01/2022. Although the transformer operates on sequences of length 100, we do not restrict each user to their most recent 100 interactions. Instead, for each user we partition the full interaction history into contiguous segments of 100 (item, action) events. Each segment is treated as an independent training example and is prepended with a user identifier token, following (\cite{liu2024multi}). This design allows the model to capture long-term user preferences and personalize predictions beyond a fixed 100-interaction window. We train the model for 50 epochs using the Adam optimizer with a learning rate of $5\times10^{-4}$ in the first training stage and reduce it to $1\times10^{-4}$ for all subsequent time windows.

\input{figures/results_representation}
\input{figures/results_ref_size}

Continual adaptation proceeds in 6-month intervals starting at 01/01/2022, which we adopt instead of single-month intervals due to the larger performance degradations observed at this timescale (see Figure~\ref{fig:setup} in Section~\ref{sec:introduction}). For each interval $I_t$, we define $D_{train}^t$ as the set of newly observed user interaction chunks of length 100 that occur within the interval and were not included in any previous training window. From $D_{train}^t$, we construct a reference set $D_{ref}^t$ by uniformly sampling 100 user histories from the most recent month in $I_t$, corresponding to the latest data available at selection time; increasing the size of $D_{ref}^t$ did not yield substantial performance differences (see Figure~5). Each data selection method described in Section~4 uses $D_{ref}^t$ to score and select a subset $D_{select}^t \subset D_{train}^t$, where the subset size is specified as a fixed fraction of the newly observed data (e.g., 20\% or 50\%). Starting from the checkpoint at the beginning of $I_t$, we then continue training the model by appending $D_{select}^t$ to the existing training data and optimizing for an additional 50 epochs with a reduced learning rate. The resulting model is evaluated on user histories from the subsequent 6-month interval $I_{t+1}$ using NDCG@10, NDCG@50, HR@10, and HR@50 for next-item prediction. For KNN-Weighted, we use $C = 10$ clusters, and for Diverse-Weighted, we use a batch size of $|B_t| = 128$.

%% file: figures/results_representation.tex
\begin{figure}[t]
    \centering
    \includegraphics[width=\textwidth]{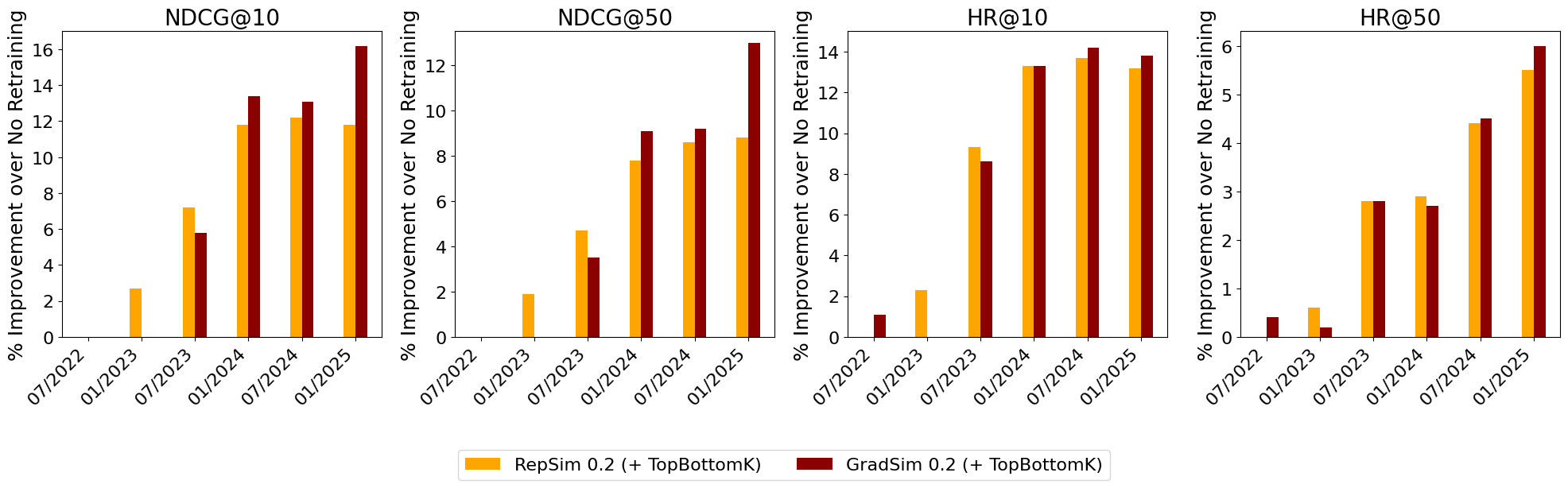}
    \caption{Comparison analysis of GradSim and RepSim representation types. Performance is measured as improvement relative to the no-retraining baseline.}
    \label{fig:results_representation}
\end{figure}

%% file: figures/results_ref_size.tex
\begin{figure}[t]
    \centering
    \includegraphics[width=\linewidth]{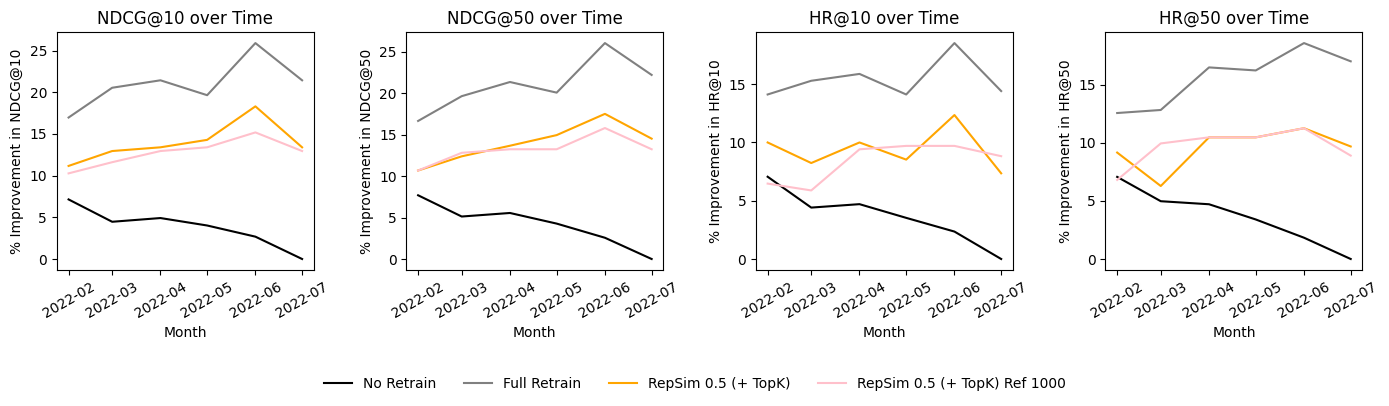}
    \caption{
    Impact of the reference set size used for RepSim selection.
    The figure reports relative performance trends for NDCG@10, NDCG@50, HR@10, and HR@50 when using reference sets of size 100 and 1000.
    All metrics are shown over time and normalized to highlight relative improvements.
    }
    \label{fig:results_ref_size}
\end{figure}

%% file: sections/results.tex
\section{Results}\label{sec:results}

\input{figures/results_topbottom_both}

\input{figures/results_topbottom_weighted}

In this continual adaptation setting, we first compare representation and sampling strategies for data selection, with gradient-based representations and diversity-aware sampling emerging as the strongest combination (Figures 4-7). We then evaluate how and when this approach improves over random data selection (Figure 8 and Table \ref{tab:error_reduction}). Additional comparisons are provided in Appendix C.

\input{tables/main_results}

\textbf{Representation Results:} Figure~\ref{fig:results_representation} compares model-based (RepSim) and gradient-based (GradSim) representations for scoring user histories during data selection. Across all evaluation metrics (NDCG and HR), GradSim consistently yields stronger downstream performance, indicating that gradient-informed signals better identify training examples that support effective adaptation under temporal drift. We omit BM25 from the main plots due to its substantially weaker performance in preliminary experiments (Appendix~C). From a systems perspective, GradSim requires both forward and backward passes to compute per-example gradients, whereas RepSim relies only on forward passes and can be nearly cost-free if representations are logged during training. Overall, these results highlight a clear trade-off between effectiveness and computational cost, with gradient-based representations providing the strongest gains, as shown in Figure ~\ref{fig:flops}. We provide additional computational analysis on these data selection methods in Appendix ~\ref{apdx:flops}.

\textbf{Reference set size:} Figure~5 shows that increasing the reference set size $|D_{ref}|$ from 100 to 1000 does not lead to meaningful changes in downstream performance. This suggests that effective distribution matching can be achieved using a small, recent reference slice, enabling scalable data selection and aligning with our emphasis on continual adaptation under compute constraints.

\input{figures/flops}

\textbf{Sampling strategy results:} Figures \ref{fig:results_topbottom_both}- \ref{fig:results_diversity} compare ranking-based and probabilistic sampling strategies. Figure \ref{fig:results_topbottom_both} shows that selecting a mixture of high-scoring and low-scoring examples (TopBottomK) outperforms selecting only the highest-scoring examples (TopK), consistent with a trade-off between matching the recent reference distribution and maintaining coverage of rarer contexts. Figure \ref{fig:results_topbottom_weighted} further shows that weighted sampling improves over hard thresholding, suggesting that preserving some mass in the middle of the score distribution can stabilize updates when scores are noisy. Finally, Figure \ref{fig:results_diversity} shows that adding explicit diversity structure (KNN-Weighted and Diverse-Weighted) further improves performance, particularly on NDCG, highlighting the benefit of combining distribution matching with diversity when adapting to drift.

To better contextualize performance under temporal drift, Table~\ref{tab:error_reduction}
reports results in terms of \emph{error reduction}, measuring how much of the performance degradation
relative to full retraining is recovered by each method \footnote{This differs from Figures 4-8 which are relative to the "no retraining" baseline, rather than the gap.}. While random subsampling already recovers a
substantial fraction of the drift-induced error, representation-aware selection is consistently more
effective. RepSim recovers over half of the lost performance at both one- and three-year horizons,
outperforming random selection at the same update budget, while gradient-based representations
achieve the strongest results overall, recovering up to 78\% of the error at longer horizons. Together,
these results show that carefully curated subsets of recent data can recover a large portion of the
benefits of full retraining, with gradient-informed selection providing the most robust improvements
under sustained drift.

\input{figures/results_diversity}

%% file: figures/results_topbottom_both.tex
\begin{figure}[t]
    \centering
    \includegraphics[width=\textwidth]{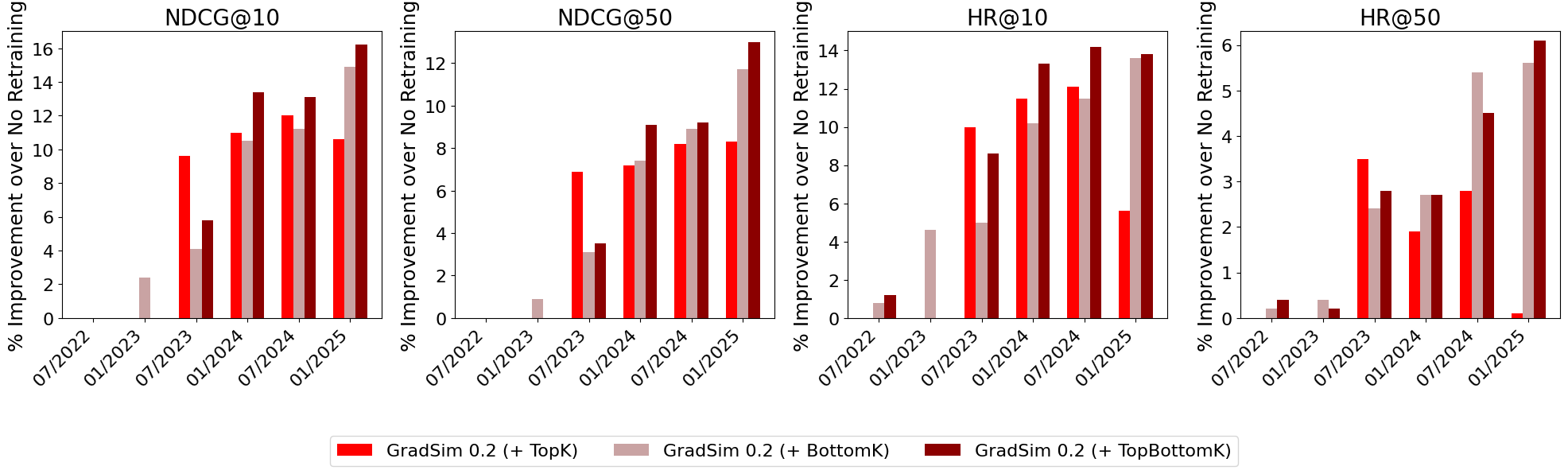}
    \caption{Comparison analysis of ranking-based sampling: top-k, bottom-k, both. Performance is measured as improvement relative to the no-retraining baseline.}
    \label{fig:results_topbottom_both}
\end{figure}

%% file: figures/results_topbottom_weighted.tex
\begin{figure}[t]
    \centering
    \includegraphics[width=\textwidth]{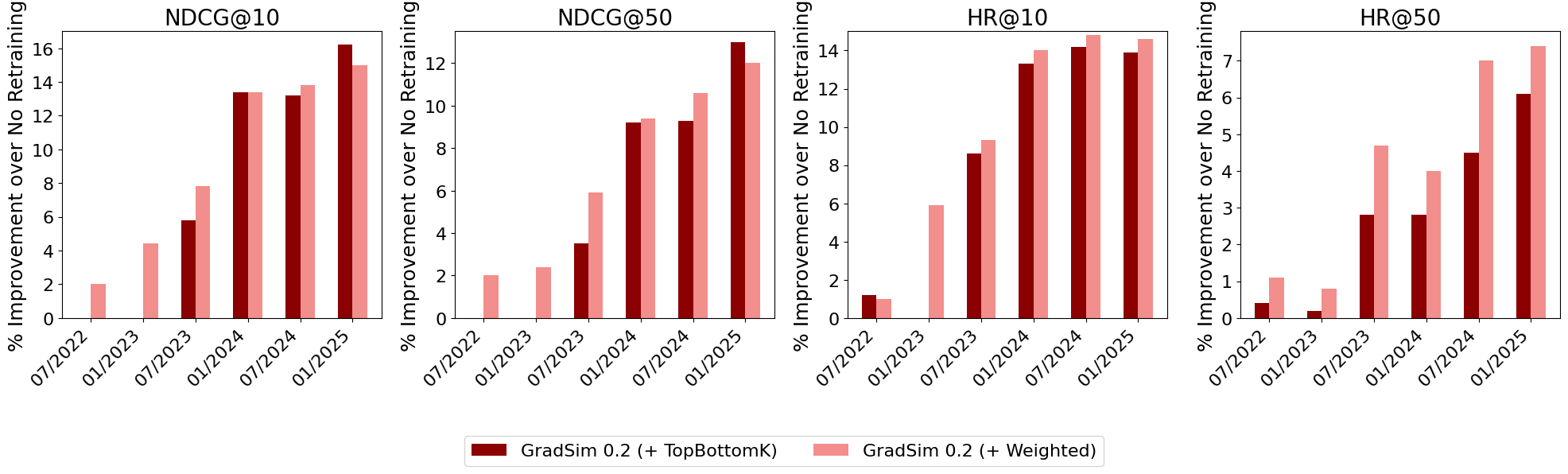}
    \caption{Comparison analysis of TopBottomk versus weighted sampling. Performance is measured as improvement relative to the no-retraining baseline.}
    \label{fig:results_topbottom_weighted}
\end{figure}

%% file: tables/main_results.tex
\begin{table}[h]
\centering
\small
\setlength{\tabcolsep}{10pt}
\begin{tabular}{lcc}
\toprule
\textbf{Method} & \textbf{+1 yr} & \textbf{+3 yrs} \\
\midrule
No Retrain & 0\% & 0\% \\
Random (20\%) & 38\% & 42\% \\
RepSim (TopBottomK) & \textbf{55\%} & \textbf{61\%} \\
GradSim (Diverse-Weighted) & \textbf{72\%} & \textbf{78\%} \\
\bottomrule
\end{tabular}
\caption{Fraction of drift-induced NDCG@50 error recovered relative to no retraining.
Error reduction is computed with respect to the gap between the no-retraining model and full retraining
at each evaluation horizon.}
\label{tab:error_reduction}
\end{table}

%% file: figures/flops.tex
\begin{wrapfigure}{r}{0.55\textwidth}
    \centering
    \begin{subcaptionbox}{Year 1 \label{fig:train}}[0.25\textwidth]
        {\includegraphics[width=\linewidth]{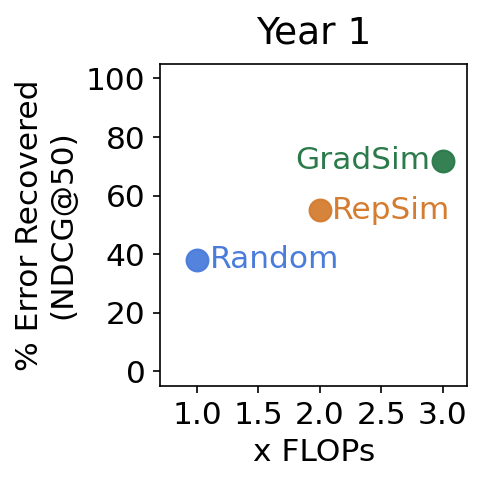}}
    \end{subcaptionbox}
    \begin{subcaptionbox}{Year 2 \label{fig_eval}}[0.25\textwidth]
        {\includegraphics[width=\linewidth]{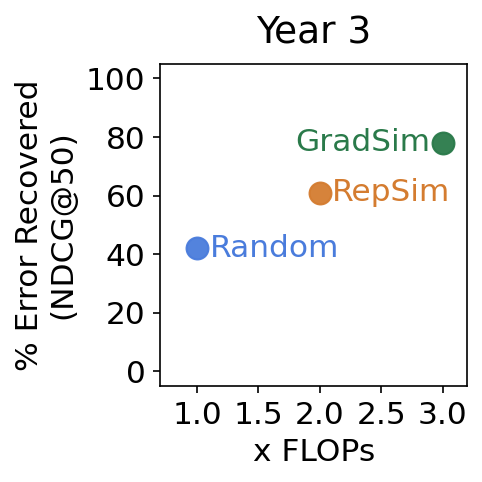}}
    \end{subcaptionbox}
    \caption{FLOPs versus performance trade-off.}
    \label{fig:flops}
    \vspace{-0.5cm}
\end{wrapfigure}

%% file: figures/results_diversity.tex
\begin{figure}[t]
    \centering
    \includegraphics[width=\textwidth]{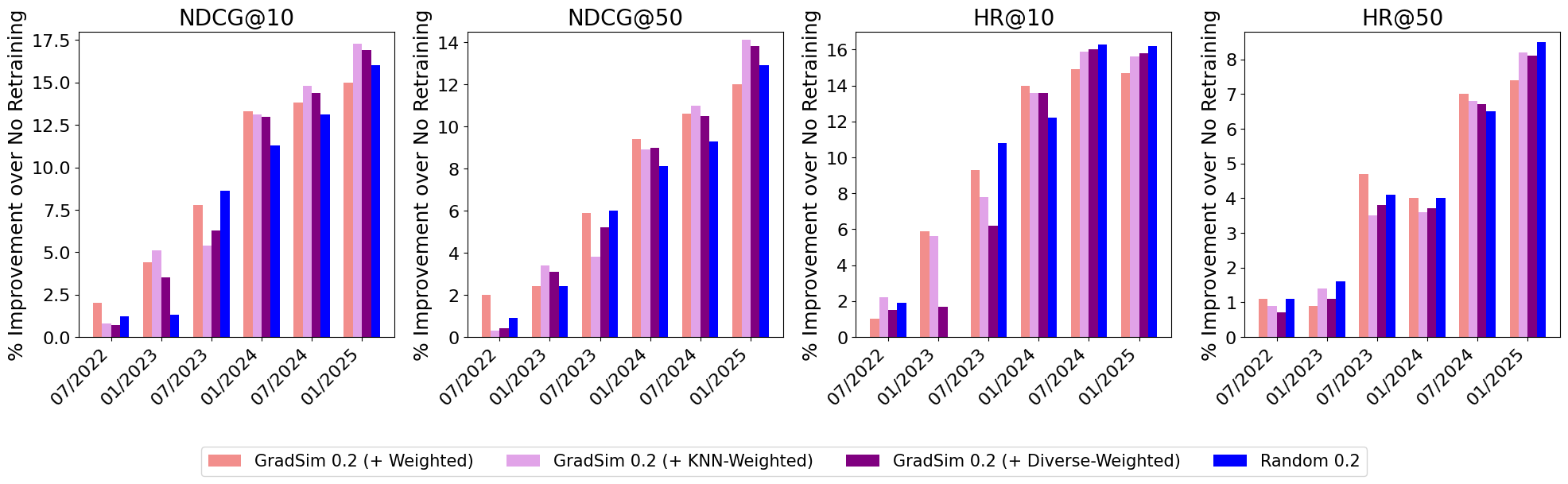}
    \caption{Comparison analysis of different sampling strategies (weighted, knn-weighted, diverse-weighted) against random selection. Performance is measured as improvement relative to the no-retraining baseline.}
    \label{fig:results_diversity}
\end{figure}

%% file: sections/conclusion.tex
\section{Conclusion and Future Work}
In this work, we study data selection for continual adaptation by examining both representation choices and sampling strategies. We find that gradient-based representations outperform model-based embeddings, accounting for a large fraction of the gains from data selection under temporal drift. On the sampling side, strategies that combine distribution matching with explicit diversity further improve performance over purely score-driven or random baselines. Together, these results show that carefully curated subsets of recent data can recover a substantial portion of the benefits of full retraining while remaining compatible with compute constraints.

Future work includes exploring representations beyond the final layer and more systematically studying interactions between representations and sampling strategies. In particular, pairing lower-cost model-based embeddings with probabilistic and diversity-aware sampling may help close the gap to gradient-informed selection. Finally, practical deployment motivates lightweight signals for detecting temporal drift and guiding when targeted data curation is necessary beyond random sampling.


%% file: sections/appendix.tex
\newpage
\section*{Appendix}
\input{sections/papers_related_work}
\input{sections/preliminary_experiments}

\input{sections/flops}

%% file: sections/papers_related_work.tex
\section{Related Work}

This appendix summarizes additional works related to data selection, temporal data attribution, and user history modeling that provide background and context for the main paper.

\subsection{Data Selection}

\textbf{Model-aware data attribution.}
A line of work uses data attribution methods to estimate the influence of individual training examples on model behavior, enabling model-aware data selection. Methods such as TracIn \citep{pruthi2020estimating} perform local probing by computing influence scores at intermediate checkpoints, while more recent approaches like MATES and LESS \citep{yu2024mates, xia2024less} adapt these ideas to large language models and improve scalability. These methods typically interleave influence estimation, data selection, and continued training, allowing selection decisions to evolve with the model.

\textbf{Model-agnostic data selection.}
In contrast, model-agnostic approaches score data without relying on gradients or model internals. DSIR \citep{xie2023data} formulates data selection as a submodular optimization problem over engineered features (e.g., token distributions), offering greater flexibility and substantially lower computational cost. Such methods enable efficient scoring at scale and provide a complementary alternative to influence-based techniques.

\textbf{Benchmarks and evaluations.}
Several works focus on benchmarking data selection and attribution methods. DCLM \citep{li2024datacomp} introduces a large-scale benchmark for data curation in language model pre-training, while DATE-LM \citep{jiao2025datelm} evaluates data attribution methods across multiple learning settings, highlighting their strengths and limitations under different objectives.

\textbf{Temporal data attribution.}
A smaller body of work studies data attribution under temporal dynamics, where data influence may change over time. ALinFiK  analyzes differences between current and future influence estimates, and related work on temporal dependence of training data influence \citep{wangcapturing} investigates how influence evolves across training horizons. These studies motivate considering time-aware data selection in non-stationary settings.

\subsection{Modeling User History}

\textbf{Sequential user modeling.}
User behavior modeling is commonly framed as a sequence modeling problem. Architectures such as SASRec \citep{kang2018self} and HSTU \citep{zhai2024actions} use self-attention to capture temporal dependencies in user interaction histories, forming the backbone of many modern recommender systems.

\textbf{User history context and compression.}
Recent production systems emphasize efficient representation of long and evolving user histories. Sliding-window approaches \citep{joshi2024sliding}, pin-based or memory-based representations \citep{pancha2022pinnerformer, badrinath2025pinrec}, and large-scale deployments at Instagram \citep{lyu2025dv365} and LinkedIn \citep{hertel2024efficient} highlight practical strategies for balancing recency, diversity, and computational constraints.

\textbf{Continuous learning over time.}
Continuous learning research addresses when and how models should be updated as data distributions shift. Prior work formulates retraining as a cost–performance tradeoff \citep{florence2025retrainmachinelearningmodel}, studies online continual learning under progressive distribution shift \citep{zhai2023online}, and introduces temporal benchmarks such as TemporalWiki and related datasets \citep{ yao2022wild}. While many benchmarks focus on knowledge retention in language models, they provide useful context for evaluating adaptation under non-stationarity.

%% file: sections/preliminary_experiments.tex
\section{Additional Experiments}\label{apdx:preliminary_experiments}

This appendix provides supporting analyses for the main paper. We first quantify the extent of temporal drift induced by a strict temporal train–test split (Section C.1). We then report baseline data selection results under fixed subset budgets and multiple evaluation horizons (Section C.2), elaborating on comparisons between random subsampling and alternative selection methods that contextualize the main paper’s focus on gradient-based representations and structured sampling strategies.

\subsection{Temporal Drift and Performance Degradation}

To validate the temporal evaluation setting, we split the dataset into training and test sets such that
all interactions prior to 01/01/2022 are used for training and all subsequent interactions are held out
for evaluation. 

As shown in Table 1, performance degrades steadily as evaluation moves further into the future, with substantial drops by one and three years. This confirms the presence of temporal drift and motivates continual adaptation rather than one-time training.

\input{tables/time_analysis}

\subsection{Baseline Selection Methods}

We next evaluate baseline data selection methods under the same temporal split to understand how subset size and simple similarity-based selection affect downstream performance. For each method, we select a fixed fraction of the pre-cutoff training data (10\%, 20\%, or 50\%). Random baselines use uniform sampling without replacement, while RepSim uses the Top-K/Bottom-K. strategy detailed in Section \ref{sec:methods}.

\input{tables/selection_1week}

At short horizons (Table \ref{tab:random_selection_relative}), selection quality matters most at smaller budgets. RepSim consistently outperforms random sampling at 10–20\%, while BM25 degrades sharply, indicating that token-based retrieval is a poor proxy for selecting effective training sequences in this setting.

\input{tables/selection_1year}

At longer horizons (Table \ref{tab:selection_1year_relative}), broader coverage becomes more important: random sampling performs strongly at small budgets, while RepSim remains competitive at higher budgets. Across all horizons, BM25 is the least effective method.

%% file: tables/time_analysis.tex
\begin{table}[h!]
\caption{Relative performance of the HSTU model on item prediction for future time periods, normalized by the initial evaluation (baseline = 1.00).}
\label{tab:test_time_relative}
\resizebox{\textwidth}{!}{
\begin{tabular}{l|ccccc}
    \toprule
    & NDCG@10 & NDCG@50 & HR@10 & HR@50 & MRR \\
    \midrule
    Initial Performance      & \textbf{1.000} & \textbf{1.000} & \textbf{1.000} & \textbf{1.000} & \textbf{1.000} \\
    Test Set (1 week)        & 0.970 & 0.970 & 0.970 & 0.970 & 0.970 \\
    Test Set (1 month)       & 0.919 & 0.921 & 0.943 & 0.922 & 0.919 \\
    Test Set (1 year)        & 0.882 & 0.882 & 0.903 & 0.893 & 0.869 \\
    Test Set (3 years)       & 0.691 & 0.689 & 0.630 & 0.630 & 0.728 \\
    \bottomrule
\end{tabular}
}
\end{table}

%% file: tables/selection_1week.tex
\begin{table}[h!]
\caption{Relative performance of the HSTU model on item prediction up to \textbf{one week} after 01/01/2022, normalized by the Full Train Set (baseline = 1.00).}
\label{tab:random_selection_relative}
\resizebox{\textwidth}{!}{
\begin{tabular}{l|ccccc}
    \toprule
    & NDCG@10 & NDCG@50 & HR@10 & HR@50 & MRR \\
    \midrule
    Full Train Set & \textbf{1.000} & \textbf{1.000} & \textbf{1.000} & \textbf{1.000} & \textbf{1.000} \\
    \midrule
    Random (50\%) & 0.922 & 0.928 & 0.950 & 0.960 & 0.910 \\
    Random (20\%) & 0.829 & 0.850 & 0.909 & 0.954 & 0.796 \\
    Random (10\%) & 0.700 & 0.762 & 0.770 & 0.936 & 0.679 \\
    \midrule
    BM25 (50\%) & 0.913 & 0.923 & 0.943 & 0.964 & 0.898 \\
    BM25 (20\%) & 0.655 & 0.702 & 0.861 & 0.954 & 0.570 \\
    BM25 (10\%) & 0.521 & 0.597 & 0.765 & 0.939 & 0.420 \\
    \midrule
    RepSim (50\%) & 0.930 & 0.936 & 0.950 & 0.962 & 0.922 \\
    RepSim (20\%) & 0.867 & 0.889 & 0.900 & 0.954 & 0.857 \\
    RepSim (10\%) & 0.769 & 0.824 & 0.798 & 0.938 & 0.774 \\
    \bottomrule
\end{tabular}
}
\end{table}

%% file: tables/selection_1year.tex
\begin{table}[h!]
\caption{Relative performance of the HSTU model on item prediction up to \textbf{one year} after 01/01/2022, normalized by the Full Train Set (baseline = 1.00).}
\label{tab:selection_1year_relative}
\resizebox{\textwidth}{!}{
\begin{tabular}{l|ccccc}
    \toprule
    & NDCG@10 & NDCG@50 & HR@10 & HR@50 & MRR \\
    \midrule
    Full Train Set & \textbf{1.000} & \textbf{1.000} & \textbf{1.000} & \textbf{1.000} & \textbf{1.000} \\ 
    \midrule
    Random (50\%) & 0.985 & 0.985 & 0.979 & 0.980 & 0.991 \\ 
    Random (20\%) & 0.960 & 0.961 & 0.977 & 0.979 & 0.955 \\ 
    Random (10\%) & 0.907 & 0.933 & 0.890 & 0.965 & 0.924 \\ 
    \midrule
    BM25 (50\%) & 0.983 & 0.978 & 0.996 & 0.980 & 0.976 \\ 
    BM25 (20\%) & 0.767 & 0.792 & 0.935 & 0.976 & 0.691 \\ 
    BM25 (10\%) & 0.708 & 0.753 & 0.858 & 0.960 & 0.644 \\ 
    \midrule
    RepSim (50\%) & 1.001 & 1.000 & 0.977 & 0.979 & 1.015 \\ 
    RepSim (20\%) & 0.912 & 0.934 & 0.922 & 0.971 & 0.919 \\ 
    RepSim (10\%) & 0.806 & 0.858 & 0.831 & 0.961 & 0.811 \\ 
    \bottomrule
\end{tabular}
}
\end{table}

%% file: sections/flops.tex
\section{Efficiency Analysis}\label{apdx:flops}
In this section, we conduct efficiency (FLOPs) analysis on the data selection methods presented in our work. Recall that given a set of training samples as $\mathcal{D}_{train} = \{x_i\}_{i=1}^n$, our objective is to select a small high-quality subset $\mathcal{D}_{select} \subset \mathcal{D}_{train}$ where $|\mathcal{D}_{select}|=k$, often with the aid of a reference set $|\mathcal{D}_{\mathrm{ref}}|=r$. We define a few variables:
\begin{itemize}
  \item $F_\text{fwd}$: FLOPs for one forward pass through $M$
  \item $F_\text{bwd} \approx 2 F_\text{fwd}$: FLOPs for one backward pass
  \item $d_\text{rep}$, $d_\text{grad}$: dimensionality of representation and gradient vectors
  \item $F_\text{train}(k)$: FLOPs to train on $k$ samples.
\end{itemize}

Given these definitions we examine the computation required for each data selection method in a single time period.

\paragraph{Random (baseline).}
No selection computation is required:
\[
  F_\text{select}^{\text{rand}} = 0
\]

\paragraph{RepSim.}
Obtain representations via a forward pass for all $n + r$ samples in $\mathcal{D}_{train}$ and $\mathcal{D}_{ref}$, then compute
pairwise cosine similarities:
\[
  F_\text{select}^{\text{rep}} = \underbrace{(n + r)\,F_\text{fwd}}_{\text{representations}}
  + \underbrace{n \cdot r \cdot d_\text{rep}}_{\text{similarity}} + \underbrace{n \log n}_{\text{top-}k\text{ sort}}
\]

\paragraph{GradSim.}
Obtain gradient representations via a forward and backward pass, then compute
pairwise similarities over (projected) gradient vectors of dimension $d_\text{grad}$:
\[
  F_\text{select}^{\text{grad}} = \underbrace{(n + r)(F_\text{fwd} + F_\text{bwd})}_{\text{gradient representations}}
  + \underbrace{n \cdot r \cdot d_\text{grad}}_{\text{similarity}} + \underbrace{n \log n}_{\text{top-}k\text{ sort}}
\]

Since $F_\text{bwd} \approx 2F_\text{fwd}$, we have
$F_\text{select}^{\text{grad}} \approx 3\,F_\text{select}^{\text{rep}}$. Table~\ref{tab:flops} summarises approximate relative FLOPs with Random $= 0$

\begin{table}[h]
\centering
\begin{tabular}{lcc}
\toprule
\textbf{Method} & \textbf{Relative FLOPs} & \textbf{Dominant cost} \\
\midrule
Random     & 0 & --- \\
RepSim     & $\approx 1$ & $+ (n+r) F_\text{fwd}$ \\
GradSim    & $\approx 3$ & $+ (n+r)(F_\text{fwd} + F_\text{bwd})$ \\
\bottomrule
\end{tabular}
\caption{Relative FLOPs per period (Random $= 0$).}
\label{tab:flops}
\end{table}

